\documentclass{elsart4-1}

\usepackage{graphicx}
\usepackage{amssymb}
\usepackage[dvips]{color}

\begin{document}

\newcommand{\vspfigA}{\vspace{1cm}}  
\newcommand{\vspfigB}{\vspace{0cm}} 
\newcommand{\vspfigC}{\vspace{0.3cm}}
\newcommand{\widthfigA}{0.6\textwidth}
\newcommand{\widthfigB}{0.7\textwidth}
\newcommand{\widthfigC}{0.7\textwidth}

\newcommand{\bfGamma}{\mbox{\boldmath $\bf\Gamma$}}


%
\begin{frontmatter}
\title{Lyapunov Modes for a Nonequilibrium System 
with a Heat Flux}

\author{ Tooru Taniguchi$^{1}$ and Gary P. Morriss$^{2}$}
%
 
%
\address{ 
${}^{1}$The Rockefeller University, 1230 York 
Avenue, New York, NY 10021, USA. \\
${}^{2}$  
School of Physics, University of New South Wales, 
Sydney, New South Wales 2052, Australia. 
} 
%


%
\date{\today}

\begin{abstract}

   We present the first numerical observation of Lyapunov modes  
(mode structure of Lyapunov vectors)
in a system maintained in a nonequilibrium steady state. 
   The modes show some similarities and some differences when 
compared with the results for equilibrium systems. 
   The breaking of energy conservation removes a zero exponent and
introduces a new mode.
   The transverse modes are only weakly altered but there are systematic
changes to the longitudinal and momentum dependent modes.

\end{abstract}

\end{frontmatter}

\section{Introduction}

   The difference of two trajectories 
starting from infinitesimally nearby initial conditions, 
which is called the Lyapunov vector, 
plays an essential role in the description of  
stability or instability in dynamical systems.  
   The exponential rate of expansion or contraction of  
the absolute value of the Lyapunov vector is  
the Lyapunov exponent, and its positivity means that  
the system has a dynamical 
instability and is called chaotic. 
   The Lyapunov vector is introduced in 
each independent direction of phase space, 
so in general we have to consider a set of 
Lyapunov exponents,  called the Lyapunov spectrum, 
and the corresponding Lyapunov vectors 
in a high-dimensional chaotic system. 
   An algorithm to calculate the Lyapunov exponents
and vectors in many-body systems 
has been developed by Benettin et al.
\cite{Ben76,Ben80a,Ben80b}, 
and Shimada and Nagashima \cite{Shi79}, 
and the behavior of Lyapunov vectors has been investigated 
from various points of view, for example,  
the conjugate pairing rule for Lyapunov spectra 
in some thermodynamic systems \cite{Dre88,ECM90,DM96,TM02}, 
and the localization behavior of Lyapunov vectors 
\cite{Man85,Kan86,Liv89,Fal91,Gia91,Mil02,TM03b,TM04b}, etc. 

   Recently, wavelike structure of Lyapunov vectors, called 
Lyapunov modes, and the corresponding stepwise 
structure of Lyapunov spectra has drawn attention in 
many-particle chaotic systems  
\cite{Del96,Pos00,Eck00,TM02c,Mcn01b,Wij04,TM02b}. 
   This structure appears in the region of  
Lyapunov exponents with small absolute values. 
   Lyapunov exponents have the dimension of inverse time, 
so the structure in the small Lyapunov exponents is supposed 
to be a reflection of the slow and macroscopic behavior 
of many-particle systems.    
   Since the first observation of Lyapunov modes in computer simulations
of hard disk systems  
\cite{Pos00,Yan04,TM03a,TM05c,Eck05}
they have been found in systems with soft potentials, independent
of dimensionality and system geometry.
   It is now believed that the zero modes arise due to the presence
of conserved quantities \cite{TM05a}, and the spatial dependence of the non-zero 
modes is essentially analogous 
to higher order Fourier components. 
   It has been shown that the breaking of conserved quantities, 
by for example a change from periodic to hard wall boundary 
conditions, leads to the absence of particular modes.    
   It is also known that there are stationary modes 
and time-oscillating modes among the non-zero Lyapunov modes, 
and the time-oscillating period of the Lyapunov modes 
is connected to that of an equilibrium momentum auto-correlation 
function \cite{TM05c,TM05a}.

   In this paper we consider the effect of breaking energy 
conservation by adding energy at one end of the system 
and removing it at the other.
   Although this leads to a flow of heat energy through the 
system, this is not intended to be a realistic model
of low dimensional heat flow.  
   Quasi-one-dimensional systems have been extensively used to 
investigate Lyapunov modes and the localization or delocalization of 
Lyapunov vectors for many-particle systems at equilibrium  
\cite{TM03a,TM05c,TM03b,TM04b}, and these systems can
easily be adapted to maintain a heat flow. 
   Using such a system, we investigate nonequilibrium effects 
of the Lyapunov modes 
and the momentum auto-correlation functions.

\section{Quasi-One-Dimensional Heat Model} 

\begin{figure}[!htbp]
\vspfigA
\begin{center}
\includegraphics[width=\widthfigA]{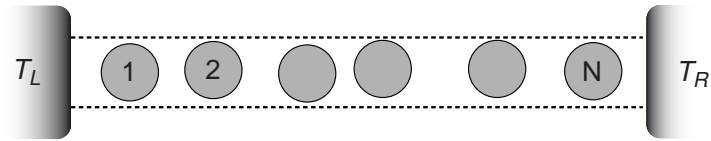}
\vspfigB
\caption{
      The quasi-one-dimensional hard-disk system with a heat flux.
      The system width is narrow so that the disks remain 
   in the same order (numbered $1,2,\cdots,N$ from left to right). 
      When disks collide with walls at either end of the system 
   energy is transferred between the disks and walls depending
   upon the values of the temperatures $T_{L}$ and $T_{R}$.  
   }
\label{figB1qua1dim}
\end{center}
\vspfigC
\end{figure}

   We consider 
a quasi-one-dimensional many-hard-disk system 
with a mechanism to insert energy at one end of
the system and  to remove energy from 
the other end. 
   It consists of $N$ hard disks of radius $R$ and 
mass $m$ in a two-dimensional rectangular region 
of length $L_{x}$ and width $L_{y}$, 
with $L_{y}< 4R$ so that the disk order remains invariant. 
    We choose periodic boundary conditions 
in the transverse $y$-direction, and
in the longitudinal $x$-direction
we use a special boundary condition 
that allows the transfer of energy 
between the system and walls:
\begin{eqnarray}
    p_{x}' &=& -(1-\epsilon) p_{x}  
       - \epsilon \sqrt{m k_{B}T}  \;\mbox{sgn}(p_{x}) 
       \label{BoundCondi1} \\
    p_{y}' &=& p_{y}
\label{BoundCondi2}\end{eqnarray}
where $\mbox{sgn}(x)\equiv x/|x|$, $k_{B}$ is 
Boltzmann's constant  and $\epsilon$ is a parameter 
($0\le \epsilon \le 1$). 
   In a collision, $(p_{x},p_{y})$ 
is the incoming momentum of the particle and
 $(p_{x}',p_{y}')$ is the outgoing momentum. 
   The parameter $\epsilon$ controls the strength
of the coupling to the ``heat reservoir" of temperature
$T= T_{\alpha}$ (where $\alpha$ is either $L$ or $R$). 
   When $\epsilon=1$ the incoming momentum $p_{x}$ 
is completely replaced by the mean thermal momentum 
$-\sqrt{m k_{B}T} \;\mbox{sgn}(p_{x})$ of
the reservoir and all information contained in the incoming
momentum is lost. 
   At $\epsilon=0$ the connection with the reservoir is
removed and the collision process is hard wall.
  Here we consider values of $\epsilon$ between
$0$ and $1$ so that the reservoir is coupled to the system
but the loss of information is not complete.
  We choose the temperatures of the reservoirs 
independently, so the temperature of 
the left-hand side reservoir is $T_{L}=10$, 
and the right-hand side reservoir is $T_{R}=1$. 
   A schematic illustration of this system 
is given in Fig. \ref{figB1qua1dim}. 
   For the numerical results that are shown in this paper, 
we use the values $m=1$, $N=100$, $R=1$, $L_{y}=2R(1+10^{-6})$ 
and  $L_{x}=1.5NL_{y}$.  
%

   Note that boundary condition (\ref{BoundCondi2}) 
with periodic boundary conditions in the $y$-direction 
guarantees that the total momentum 
in the $y$-direction is conserved.  
   The parameter $\epsilon$ 
determines the strength of the coupling between the
reservoir and the system, and specifies the fraction
of incoming momentum $p_{x}$ that is retained  
after the collision. 
   If $T_{L}\neq T_{R}$ and $\epsilon \neq 0$ 
then an energy transfer (heat) occurs across the system, and 
a nonequilibrium steady state is established 
after a long time.

\section{Stepwise Structure of Lyapunov Spectra}
\label{StepwiseStructureLyapunovSpectra}

\begin{figure}[!t]
\vspfigA
\begin{center}
\includegraphics[width=\widthfigA]{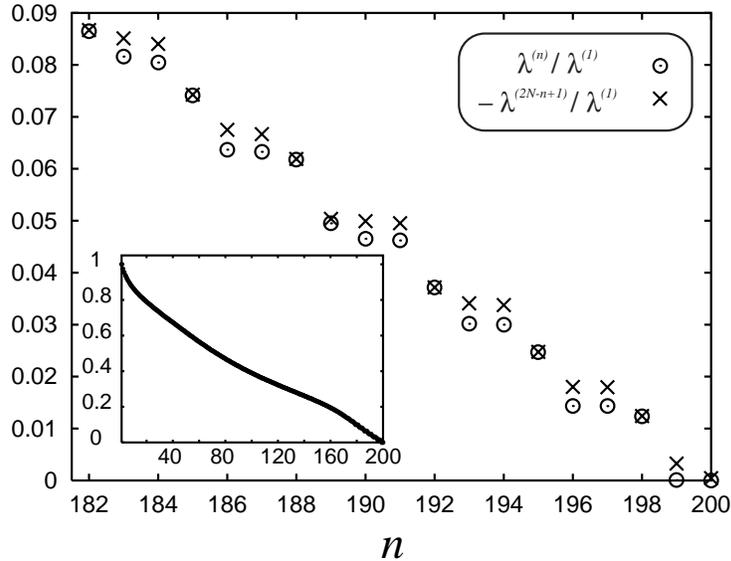}
\vspfigB
\caption{ 
      The Lyapunov spectrum for the quasi-one-dimensional 
   heat model consisting of $100$ disks with $\epsilon = 0.5$.
      The reduced Lyapunov exponents 
   ($\{|\lambda^{(n)}|/\lambda^{(1)}\}_{n}$) with smallest 
   magnitude are shown in the main plot.
      The inset is a plot of the positive exponent region of the full
   Lyapunov spectrum. 
      The circles are the smallest positive exponents, while the
   crosses are the absolute values of the conjugate 
   negative exponents. 
      The horizontal axis is the exponent number $n$, numbered
    from $1$ (the largest) to $4N=400$ (the smallest). 
   }
\label{figC1Lyapu}
\end{center}
\vspfigC
\end{figure}  
%
   Figure \ref{figC1Lyapu} is a graph of the absolute value of the 
normalized Lyapunov spectrum $\{|\lambda^{(n)}|/\lambda^{(1)}\}_{n}$
with $\epsilon=0.5$.
   Here the largest Lyapunov exponent 
is $\lambda^{(1)}\approx 1.79$.  
   From this figure we see that the positive 
branch and the negative branch of the Lyapunov spectrum 
are not symmetric, namely $\lambda^{(j)} \neq -\lambda^{(4N-j+1)}$ 
particularly for the exponents of smallest magnitude near $N=200$. 
  This is different from the equilibrium case ($\epsilon = 0$) 
where the dynamics is Hamiltonian with the 
symplectic structure guaranteeing 
the conjugate pairing rule for the Lyapunov spectra 
$\lambda^{(j)} = -\lambda^{(4N-j+1)}$ \cite{A89}. 
   The system has 3 zero-Lyapunov exponents, 
due to total momentum conservation
and spatial translational invariance 
in the $y-$direction, and 
time-translational invariance. 
   In the nonequilibrium case ($\epsilon \neq 0$)  
the total energy of the system is no longer conserved 
and this is the reason that $\lambda^{(202)}\neq 0$, 
although it is zero at equilibrium. 
   
\begin{table}[!steps]
\vspfigA
\caption{Classification of the exponents in the Lyapunov 
spectrum for the quasi-one-dimensional system of 
$100$ particles. The four zero exponents 
are 199, 200, 201 and 202 at equilibrium.}
\begin{center}
\vspace{0.3cm}
\begin{tabular}{c|cc|cc}
    \makebox[2cm][c]{}                               & 
    \makebox[2cm][r]{Positive}                 & \makebox[2cm][l]{branch} &
    \makebox[2cm][r]{Negative}               & \makebox[2cm][l]{branch} \\
 \hline \hline
    \makebox[2cm][c]{Step number}       &
    \makebox[2cm][c]{1-point step}        & \makebox[2cm][c]{2-point step} &
    \makebox[2cm][c]{1-point step}        & \makebox[2cm][c]{2-point step} \\
 \hline
First        &  198 & 197, 196 & 203 & 204, 205\\
Second &  195 & 194, 193 & 206 & 207, 208 \\
Third      &  192 & 191, 190 & 209 & 210, 211 \\
 \hline
\end{tabular}
\end{center}
\label{default1}
\vspfigC
\end{table}%
   
   In Refs. \cite{TM03a,TM05c,TM05a}, it is shown 
that for the equivalent equilibrium system (that is, 
$\epsilon = 0$ and hard-wall boundary conditions 
in the longitudinal direction and periodic boundary 
conditions in the transverse direction), 
the Lyapunov spectrum  
consists of 1-point steps and 2-point steps. 
   The 1-point steps correspond to Lyapunov vectors 
containing transverse Lyapunov modes,  
and have their origin in total momentum 
conservation (and translational invariance) 
in the $y-$direction. 
   On the other hand, the 2-point steps 
correspond to longitudinal and momentum 
proportional mode contributions.  
   In Fig. \ref{figC1Lyapu} we can see that,
away from equilibrium,
the 1-point steps and 2-point steps remain 
(see Table \ref{default1} 
for meaning of 1-point and 2-point steps 
of the Lyapunov spectrum). 
   The positive and negative branch 1-point 
steps are symmetric, that is  
$\lambda^{(n)}\approx - \lambda^{(4N-n+1)}$, for 
$n=198,195,192,189,188,185,\cdots$. 
   This may be because total momentum conservation 
and spatial translational invariance 
in the $y-$direction is preserved
regardless of the value of $\epsilon$. 
   On the other hand, the step-heights of the 2-point steps 
are clearly different in the positive and negative branches.
   Indeed, the asymmetry of the positive and negative exponents
appears to be of the same magnitude for each 2-point step, and
as well as for exponent $202$ that has shifted from zero. 

   There is no Gaussian thermostat \cite{ECM90,DM96} 
in this system so conjugate pairing of 
Lyapunov exponents would not be expected to hold.
   The magnitudes of the largest and smallest Lyapunov
exponents appear to be almost equal, and there is only weak
evidence for phase space dimensional contraction 
because dissipation as the origin of such a contraction 
can occur only at the two particles at the ends of the system.

\section{Mode Structure of Lyapunov Vectors}

   For each Lyapunov exponent in the Lyapunov spectrum 
there is a corresponding
Lyapunov vector which will in general have both coordinate and 
momentum components.
   For the exponents of smallest magnitude, the Lyapunov vectors 
often contain either coordinates of momentum components that vary
sinusoidally with particle position forming a delocalized structure. 
   We refer to such Lyapunov vectors containing 
delocalized structures as Lyapunov modes. 

\subsection{Modification of zero-Lyapunov Modes}

\begin{figure}[!t]
\vspfigA
\begin{center}
\includegraphics[width=\widthfigB]{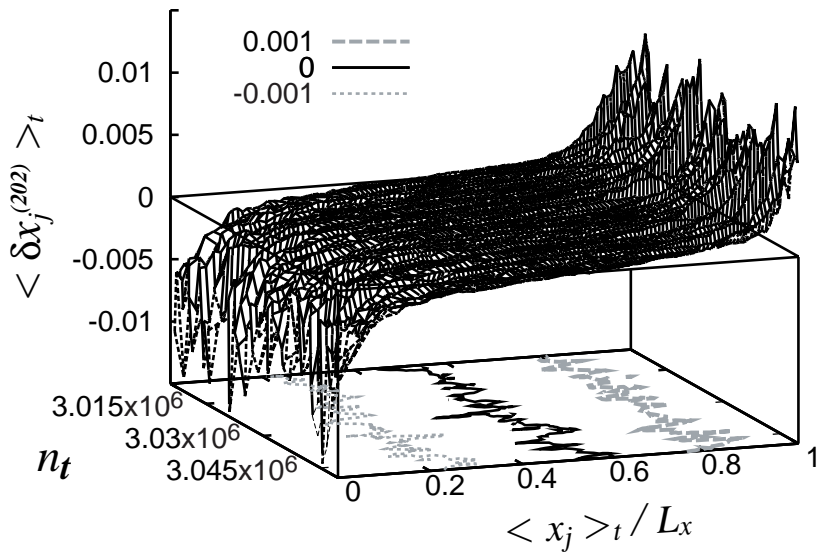}
\vspfigB
\caption{ 
      The local time average 
   $\langle\delta x_{j}^{(202)}\rangle_{t}$ 
   of the longitudinal spatial component 
   of the Lyapunov vector corresponding 
   to Lyapunov exponent $\lambda^{(202)}$ 
   as a function of the collision number $n_{t}$ 
   and the normalized local time average 
   of the particle position
   $\langle x_{j} \rangle_{t}/L_{x}$, 
   for  $\epsilon = 0.5$.  
   }
\label{figD1zeroMod}
\end{center}
\vspfigC
\end{figure}

   The introduction of the nonequilibrium boundary condition 
$\epsilon \neq 0$ leads to a change in the structure of the
Lyapunov mode corresponding to exponent $\lambda^{(202)}$. 
   The $x-$component of the new mode is shown in Fig.  
 \ref{figD1zeroMod}.
   The fact that $\lambda^{(202)} \neq 0$,  
is a purely nonequilibrium effect.  
   Figure \ref{figD1zeroMod} shows the 
longitudinal component $\delta x_{j}^{(202)}$ 
of the Lyapunov vector $\delta\bfGamma^{(202)}$ 
as a function of the collision number $n_{t}$ 
and the normalized particle position   
for  $\epsilon = 0.5$. 
   Note that in this paper 
the Lyapunov vector $\delta\bfGamma^{(n)}$ 
corresponding to the Lyapunov exponent $\lambda^{(n)}$ 
is normalized so that  $|\delta\bfGamma^{(n)}|=1$. 
   In order to obtain a clear graph, we  
take a  local time average, indicated by the notation 
$\langle\cdots\rangle_{t}$, over $8N$ collisions 
using data just after collisions. 
   Figure \ref{figD1zeroMod} shows an approximately 
steady structure with a slope which is steepest near the two 
ends of the system.  
   This structure is quite different from the equilibrium case, 
where the longitudinal component $\delta x_{j}^{(2N+2)}$ 
is found numerically to be almost zero in both space and time.

%

\subsection{Transverse Lyapunov Modes}

\begin{figure}[!t]
\vspfigA
\begin{center}
\includegraphics[width=\widthfigA]{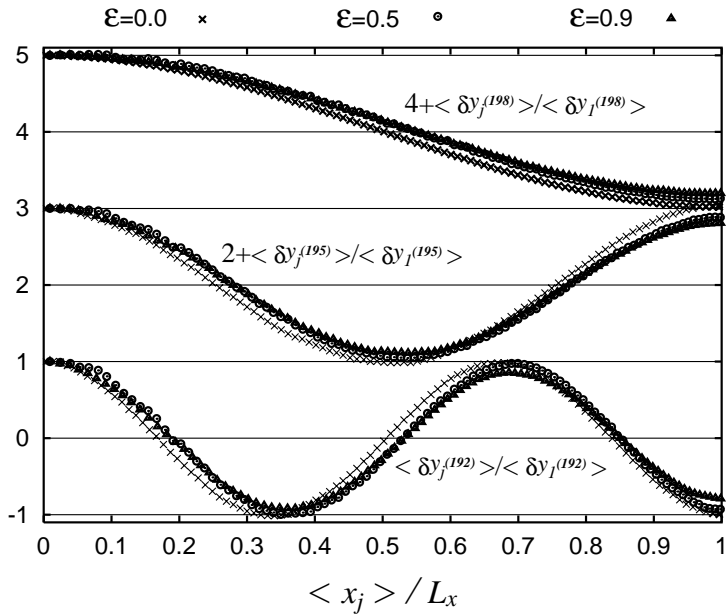}
\vspfigB
\caption{ 
      The transverse Lyapunov modes 
   for the quasi-one-dimensional heat model
   as shifted plots of the normalized 
   transverse Lyapunov vector component  
   $\langle\delta y_{j}^{(n)}\rangle
   /\langle\delta y_{1}^{(n)}\rangle$ as 
   functions of $\langle \delta x_{j}^{(n)}\rangle/L_{x}$  
   for $\epsilon =0$ (crosses), $\epsilon=0.5$ (circles) 
   and $\epsilon=0.9$ (triangles). 
      The first 1-point step ($n=198$) is shifted so that
   zero corresponds to $4$ on the vertical axis, 
   the second 1-point step ($n=195$) is shifted so that 
   zero corresponds to $2$ on the vertical axis and   
   the third 1-point step ($n=192$) is not shifted. 
      Notice that the transverse modes are largely 
   independent of the parameter $\epsilon$. 
   }
\label{figD2modeTrans}
\end{center}
\vspfigC
\end{figure}  
%
   Each 1-point step of the Lyapunov spectrum has an associated
Lyapunov vector containing a transverse mode.
   The $y-$component $\delta y_{j}^{(n)}$ for the $j$-th disk 
of the $n^{th}$ Lyapunov vector has a sinusoidal dependence
on the particle position $x_{j}/L_{x}$ 
at equilibrium \cite{TM03a,TM05c}. 
    We show that this structure is only slightly changed 
when the system is driven away from equilibrium, 
$\epsilon\neq 0$. 
   In Fig. \ref{figD2modeTrans} we present the 
time average of the transverse mode $\langle\delta y_{j}^{(n)}\rangle$ 
as functions of the average position 
$\langle \delta x_{j}^{(n)}\rangle/L_{x}$ 
of the $j$-th disk for 
different values of $\epsilon$. 
    The first ($198$), second ($195$) and third ($192$) 
transverse modes 
are clearly visible for the nonequilibrium cases.  
   Note that the transverse Lyapunov modes 
in Fig. \ref{figD2modeTrans} show 
a weak $\epsilon$-dependence. 
   In the nonequilibrium case $\epsilon\neq 0$, 
there is a deviation from a sinusoidal  curve 
in the transverse mode, while 
in the equilibrium case $\epsilon=0$, 
the mode structure can be fitted nicely by a sinusoidal curve.  
   Fig. \ref{figD2modeTrans} shows that 
the amplitude of the transverse mode decreases slightly 
from the high temperature region (left side) 
to the low temperature region (right side). 
   This change in amplitude in the transverse Lyapunov 
modes is a nonequilibrium effect .

\subsection{Longitudinal and Momentum-Proportional Lyapunov Modes}

\begin{figure}[!t]
\vspfigA
\begin{center}
\includegraphics[width=\widthfigC]{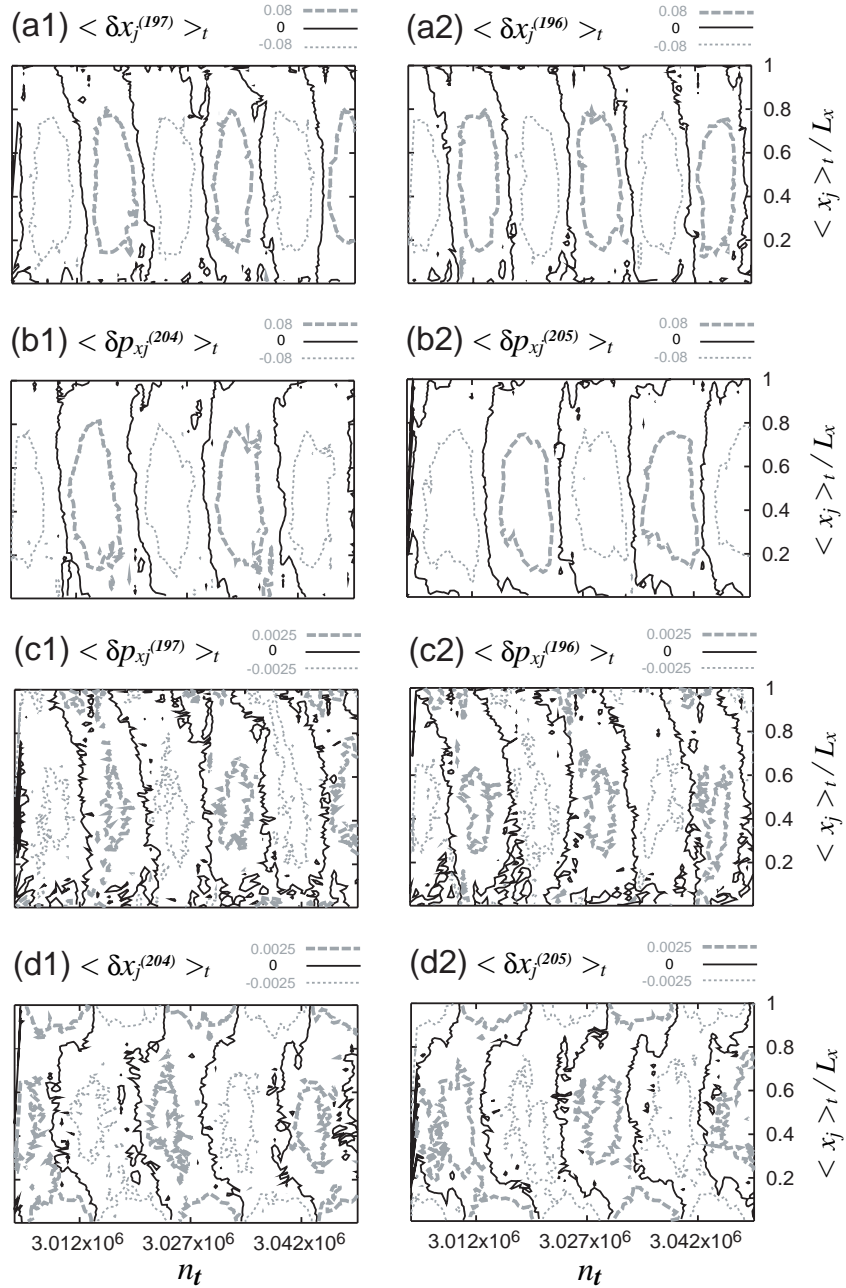}
\vspfigB
\caption{ 
      Contour plots of longitudinal Lyapunov modes 
   for the quasi-one-dimensional heat model 
   for $\epsilon=0.5$.
   Exponents $196$ and $197$ are the first 2-point 
   step in the positive branch of the spectrum and 
   $204$ and $205$ are the first 2-point step in 
   the negative branch of the spectrum.
   }
\label{figD3modLong}
\end{center}
\vspfigC
\end{figure}  

%
   Some of the structure of the longitudinal modes in the 
system remains as the system departs from equilibrium.
   There is a $\pi/2$ phase shift in time
between the $\delta x$ components of modes in the 
same 2-point step, that is between $\langle \delta x^{(197)}_{j}\rangle $ and 
$\langle \delta x^{(196)}_{j}\rangle $, 
see Fig. \ref{figD3modLong}. 
   As the momentum components are approximately the
same functional form as the spatial components, there
is the same phase shift between $\langle \delta p^{(197)}_{xj}\rangle $ 
and $\langle \delta p^{(196)}_{xj}\rangle $.
   The same behavior is observed for the components of
the modes in the first 2-point step in the negative branch
of the spectrum, $\langle \delta x^{(204)}_{j}\rangle $ and
$\langle \delta x^{(205)}_{j}\rangle $ differ by a $\pi/2$ phase shift
in time, as do the contributions $\langle \delta p_{xj}^{(204)}\rangle $ and
$\langle \delta p_{xj}^{(205)}\rangle $.
   In the positive branch 2-point step the spatial contributions
are larger than the momentum contributions, in the negative
branch the relative sizes of the contributions are reversed.

   The principle new feature is that nodal lines are curved,
with those contributions to the positive branch of the spectrum
having the
center of the system reaching the node later than the two ends.
   This gives the appearance of a ``forward'' moving wave in
the positive branch modes and a ``backward'' moving wave
in the negative branch modes.  
   It is less apparent in Fig. \ref{figD3modLong} that the 
time-oscillating period in the negative branch is larger than
that for the positive branch.  
   Also, the $\delta x$ components in the negative branch
do not have nodes at the two ends but these are shifted
within the boundaries.
%

\begin{figure}[!t]
\vspfigA
\begin{center}
\includegraphics[width=\widthfigC]{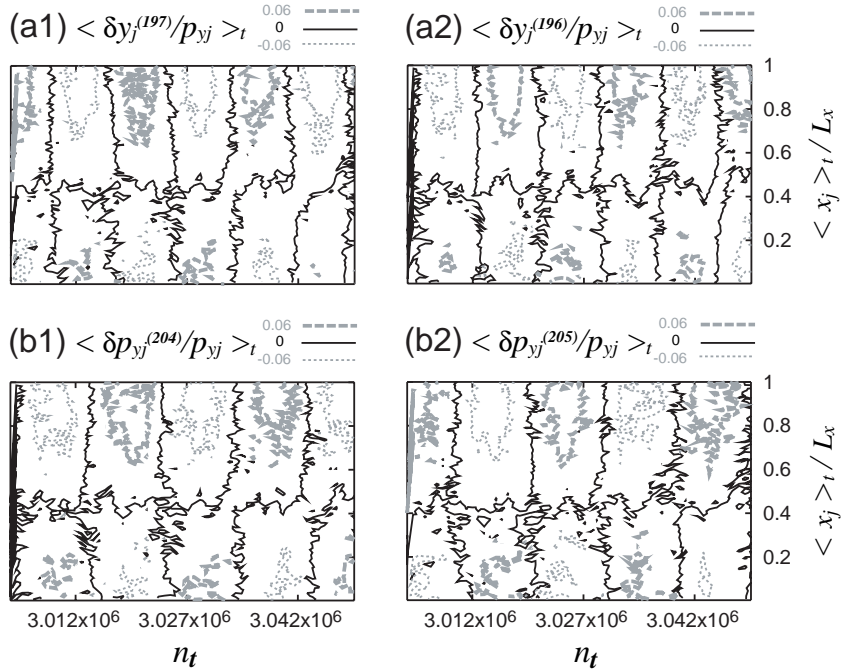}
\vspfigB
\caption{ 
      Contour plots of momentum proportional Lyapunov modes 
   for the quasi-one-dimensional heat model 
   for $\epsilon=0.5$.
      Notice that in all of these figures the nodal lines that
   would be at $\langle x_{j}\rangle_{t}/L_{x}=0.5$ for $\epsilon=0$
   are shifted to a smaller values.
   }
\label{figD4modMomen}
\end{center}
\vspfigC
\end{figure}  

   The momentum proportional components of the modes in 
Fig. \ref{figD4modMomen}  are also similar to those at equilibrium. 
   The phase differences of $\pi /2$ and the straight nodal
lines remain the same.
   However, the central nodal line at $\langle x_{j}\rangle_{t}/L_{x} = 0.5$
is now shifted towards the high temperature end of the system
for both $\langle\delta y_{j}/p_{yj}\rangle_{t}$ and $\langle\delta p_{yj}/p_{yj}\rangle_{t}$.
   As with the longitudinal modes, the time-oscillating period in
the negative branch is larger than in the positive branch.
   The other modes that are not shown in Fig. \ref{figD4modMomen} 
have a similar structure, for example 
$\langle\delta x_{j}/p_{xj}\rangle$ for $196$ and $197$ have a similar structure to 
$\langle\delta y_{j}/p_{yj}\rangle$ but with more apparent noise.
   Also, $\langle\delta p_{xj}/p_{xj}\rangle$ for $204$ and $205$ have a similar structure to 
$\langle\delta p_{yj}/p_{yj}\rangle$ but with more apparent noise.

%

\section{Time-Oscillating Periods of Lyapunov modes and Velocity Auto-Correlation}

\begin{figure}[!t]
\vspfigA
\begin{center}
\includegraphics[width=\widthfigA]{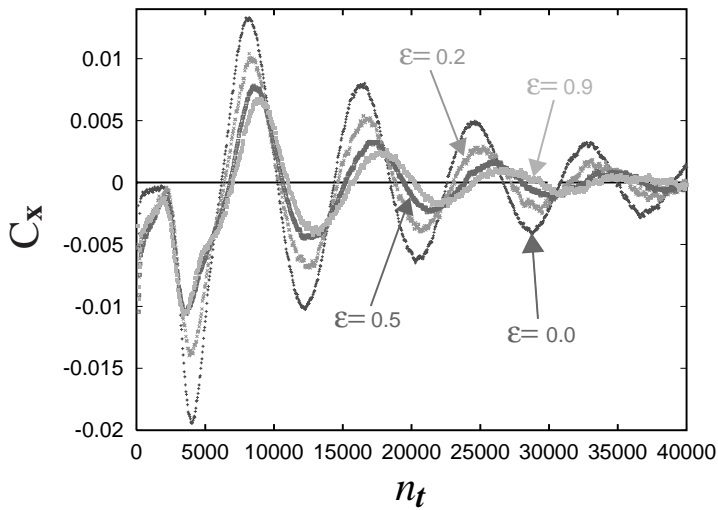}
\vspfigB
\caption{ 
      The oscillating part 
   of the normalized velocity auto-correlation function 
   $C_{x}$ as a function of the collision number $n_{t}$. 
      Here $\epsilon$ varies from $0$ to $0.9$. 
      The values of $C_{x}$ are 
   the arithmetic averages 
   of the auto-correlation for 11 disks 
   in the center of the system,      
   and is normalized 
   so that $\left.C_{x}\right|_{n_{t}=0}=1$. 
   }
\label{figE1heatACF}
\end{center}
\vspfigC
\end{figure}  

\newcommand{\ls}{\hspace{0.1\textwidth}}
\begin{table}[!thdp]
\caption{Time-oscillating period of the momentum auto-correlation
 function and Lyapunov modes}
\begin{center}
\begin{tabular}{c|ccc}
    \makebox[2cm][c]{$\epsilon$} & \makebox[2cm][c]{0} & 
    \makebox[2cm][c]{0.5}        & \makebox[2cm][c]{0.9} \\
 \hline\hline
 $T_{acf}$& 8210 \ls & 8560 \ls & 8830 \ls \\
 \hline
 $T^{(3)}$ & 5700 & 5700 & 5600 \\
 $T^{(2)}$ & 8300 & 8200 & 8100 \\
 $T^{(1)}$ & 16400 & 16200 & 14900 \\
 $T^{(-1)}$ & 16400 & 19600 & 56400 \\
 $T^{(-2)}$ & 8300 & 9300 & 12500 \\
 $T^{(-3)}$ & 5700 & 6300 & 8100 \\
 \hline
\end{tabular}
\end{center}
\label{default2}
\end{table}%

%
%
   The period $T_{acf}$ (in numbers of collisions) of the 
auto-correlation function (acf) $C_{x}$ 
for the longitudinal component of the momentum increases, 
and the amplitude decreases as a function of $\epsilon$
(see Table \ref{default2}). 
   The time oscillating period $T^{(k)}$ of 
the Lyapunov vectors for the $k$-th 
longitudinal (and the momentum proportional) 
mode in the positive branch of the Lyapunov spectrum 
decreases slightly as a function of $\epsilon$, 
whereas in the negative branch the period
increases slightly as a function of $\epsilon$. 
      Such changes of $T^{(-k)}$ are especially large 
for small $k$'s, which correspond to the ones 
close to the zero-Lyapunov exponents.  
    The relation  $T^{(1)}=2 T_{acf}$ is not correct 
away from equilibrium, while it is justified 
in equilibrium \cite{TM05c,TM05a}. 
      This does not contradict the 
proof of this relation as that required equilibrium 
and is based on the functional 
form of the Lyapunov modes, in particular that they 
are purely sinusoidal.
%

\section{Conclusion and Remarks}

   The changes in the structure of the Lyapunov modes
for a system maintained away from equilibrium by a 
boundary imposed heat flow are described in detail. 
   The breaking of conservation of energy is shown
to introduce systematic effects, changing the number
of zero exponents, and modifying the form of other 
modes, with the basic structure of transverse, longitudinal
and momentum proportional modes remaining.
   This work is largely descriptive and the theoretical
origin of the changes in modes is yet to be developed,
as it is for the equilibrium case.
    However, it is clear that the modes are not only
equilibrium properties of many-particle systems, but
also fundamental for nonequilibrium steady states.

   The original purpose in introducing the 
nonequilibrium boundary 
conditions [i.e. Eqs. (\ref{BoundCondi1}) and (\ref{BoundCondi2})] 
was to keep dynamical properties, like 
the momentum conservation in the transverse direction 
and deterministic property of orbit, etc., in 
a nonequilibrium dynamics
while breaking energy conservation. 
   The equilibrium case is recovered in the limit as 
$\epsilon \rightarrow 0$. 
   Therefore this choice of the boundary conditions 
was not intended to provide a realistic model for the 
interaction between a heat reservoir and a particle system.  
  For example, a consequence of these boundary conditions 
is that the momentum distribution function for a particle 
in contact with the reservoir 
is very different from a Gaussian.
   Indeed, we found that 
the incoming momentum is close to Gaussian 
but the outgoing momentum is strongly peaked
around the mean momentum of the reservoir,
with the width of its distribution determined by the
degree of mixing with the incoming momentum 
(that is, determined by the value of $\epsilon$).

\section*{Acknowledgement}

   The authors appreciate a careful reading 
of the manuscript by E. G. D. Cohen, and 
the financial support 
of the Japan Society for the Promotion of Science.



\end{document}